\documentclass[preprint,prd,aps,showpacs]{revtex4}
\usepackage{graphicx}
\usepackage{latexsym}
\usepackage{color}
\usepackage{amsmath}
\begin{document}
\setcounter{page}{1}
\title{Self Energy Correction in Light Front QED And Coherent State Basis}
\author{Jai D. More\footnote{\tt more.physics@gmail.com}}
\author{Anuradha Misra\footnote{\tt misra@physics.mu.ac.in}}
\affiliation{Department of Physics,University of Mumbai,
\\Santa Cruz(E), Mumbai, India-400098}%

\begin{abstract}
We discuss the calculation of fermion self energy correction in Light Front QED using a coherent state basis. We show that if one uses coherent state basis instead of fock basis to calculate the transition matrix elements, the true infrared divergences in electron mass renormalization $\delta m^2$  get canceled up to $O(e^4)$ in  
Light Front gauge. We have also verified this cancellation in Feynman gauge up to $O(e^2)$. 
\end{abstract}
\date{\today}
\maketitle
\section{INTRODUCTION}
 LSZ formalism of quantum field theory is based on the assumption that at large times, the dynamics of incoming and outgoing particles in a scattering process is governed by the free Hamiltonian, i.e. the asymptotic Hamiltonian $H_{as} $ is the same as free Hamiltonian:
\begin{equation}
H_{as}=\lim_{\left |t\right| \rightarrow \infty} H= H_0
\end{equation}

However, it was pointed out by Kulish and Faddeev \cite{KUL70} that this assumption does not hold for theories in which either 
\begin{itemize}
\item long range interactions like QED are present or 
\item the incoming and outgoing states are bound states like in QCD.
\end{itemize}
 Kulish and Faddeev (KF) proposed the method of asymptotic dynamics and showed that in QED, at large times, when one takes into account the long range interaction between the incoming and outgoing states, then 
\begin{equation}
H_{as}=H_0 +V_{as}
\end{equation}
$V_{as} $ was shown to be non-zero in QED and was used to construct the asymptotic M\"oller operators
\begin{displaymath}
\Omega_{\pm}^A = T~exp\biggl[ -i \int^0_{\mp \infty} V_{as}(t)dt \biggr]
\end{displaymath}
which leads to the coherent states
\begin{displaymath}
\vert n \colon coh \rangle = \Omega_{\pm}^A \vert{n}\rangle \;,
\end{displaymath} 
 as the asymptotic states,
where $\vert{n}\rangle$ is the n particle Fock state. 
It was then shown that the transition matrix elements evaluated between these coherent states are infra-red (IR) divergence free.\\

In this talk, I will discuss the issue of cancellation of IR divergences in the electron mass renormalization in light front QED.\\
	
For the sake of completeness, we state the notation followed by us \cite{MUS91}. Our metric tensor is 
\[
g^{\mu\nu} =
\left[{\begin{array}{cccc}
0&1&0&0\\
1&0&0&0\\
0&0&-1&0\\
0&0&0&-1\\
\end{array}}\right]
\]
so that the four vector is defined as 
 \begin{displaymath}
 x^{\mu}=(x^+,x^-,{\bf x}^{\perp}) 
\end{displaymath}
where
\begin{displaymath}
x^+=\frac{(x^0 +x^3)}{\sqrt{2}}, x^-=\frac{(x^0 -x^3)}{\sqrt{2}}, {\bf x}_{\perp}=(x^1,x^2)
\end{displaymath}
Four momentum is $p=(p^+, p^-, {\bf p_\perp})$ and the mass shell condition is  
\begin{displaymath}
p^-=\frac{p^2_\perp+m^2}{2p^+}
\end{displaymath}

In LFFT, there are two kinds of IR divergences
\begin{itemize}
\item[1)] Spurious IR divergences, which are divergences arising due to $ k^+\rightarrow 0$ and are actually a manifestation of UV divergences of equal time theory. These can be regularized by an infrared cutoff on small values of longitudinal momentum. 
\item[2)] True IR divergences are the actual IR divergences of equal time theory and are present because of the particles being on mass shell. The coherent state method is one possible way to deal with this kind of divergences.
\end{itemize}
A coherent state approach  based on the method of asymptotic dynamics  has been developed and applied to lowest order calculations in LFFT \cite{ANU94, ANU96, ANU00}. It was shown \cite{ANU94, ANU00}  that the true IR divergences do not appear when one uses these coherent states to calculate the transition matrix elements for evaluating one loop vertex correction both in LFQED and LFQCD.\\

Light-front QED Hamiltonian in the light-front gauge consists of the free part, 
the standard three point QED vertex and two 4-point instantaneous interactions. 
\begin{displaymath}
P^-= H \equiv H_0 + V_1 + V_2 + V_3 \;,
\end{displaymath}
Here
\begin{displaymath}
H_0= \int d^2 {\bf x}_\perp dx^- \{\frac{i}{2} \bar{\xi}\gamma^-
\stackrel{\leftrightarrow}{\partial}_-\xi + \frac{1}{ 2} (F_{12})^2-\frac{1}{2}a_+\partial_- \partial_k a_k \}
\end{displaymath}
\begin{displaymath}
V_1=e \int d^2{\bf x}_\perp dx^- \bar{\xi} \gamma^{\mu}\xi a_\mu\;
\end{displaymath}
\begin{displaymath}
V_2=-\frac{ie^2}{4}\int d^2{\bf x}_\perp dx^-dy^-\epsilon(x^--y^-)\\(\bar \xi a_k \gamma^k)(x)\gamma^+(a_j\gamma^j\xi)(y)
\end{displaymath}
\begin{displaymath}
V_3=-\frac{e^2}{4}\int d^2{\bf x}_\perp dx^-dy^-(\bar\xi \gamma^+\xi)(x) \vert x^--y^-\vert (\bar \xi\gamma^+\xi)(y)
\end{displaymath}
$\xi(x)$ and $a_\mu(x)$ can be expanded in terms of creation and annihilation operators as
\begin{align}
\xi(x)=&\int \frac{d^2 {\bf p}_\perp}{(2\pi)^{3/2}}\int \frac{dp^+}{\sqrt{2p^+}} \sum_{s=\pm\frac{1}{2}}
[u(p,s)e^{-i(p^+x^--{\bf p}_\perp x_\perp)} b(p,s,x^+)\nonumber\\ 
&+v(p,s)e^{i(p^+x^--{\bf p}_\perp x_\perp)}d^\dagger(p,s,x^+)],\nonumber
\end{align}
\begin{align}
a_\mu(x)=\int\frac{d^2{\bf q}_\perp}{(2\pi)^{3/2}}\int \frac{dq^+}{\sqrt{2q^+}} \sum_{\lambda=1,2}\epsilon^\lambda_\mu(q)[e^{-i(q^+x^--{\bf q}_\perp x_\perp)} a(q,\lambda,x^+)+e^{i(q^+x^--{\bf q}_\perp x_\perp)}a^\dagger(q,\lambda,x^+)],\nonumber
\end{align}
The creation and annihilation operator satisfy 
\begin{align}\label{anticommutation}
\{b(p,s),b^\dagger(p^\prime,s^\prime)\}=\delta(p^+-p^{\prime+})\delta^2({\bf 
p_\perp-p^\prime_\perp})\delta_{ss^\prime}=\{d(p,s),d^\dagger(p^\prime,s^\prime)\},
\end{align}
\begin{equation}\label{commutation}
[a(q,\lambda),a^\dagger(q^\prime,\lambda^\prime)]=\delta(q^+-q^{\prime+})\delta^2({\bf q_\perp-q^\prime_\perp}) \delta_{\lambda\lambda^\prime}.
\end{equation}

The light cone time dependence of the interaction Hamiltonian is given by 
\begin{displaymath}
H_I(x^+)=V_1(x^+)+V_2(x^+)+V_3(x^+)
\end{displaymath}
where
\begin{equation}\label{V_1as}
V_1(x^+)= e\sum_{i=1}^4 \int d\nu_i^{(1)}[ e^{-i \nu_i^{(1)} x^+} {\tilde h}_i^{(1)}(\nu_i^{(1)})+e^{i \nu_i^{(1)} x^+}{\tilde h}^{(1)\dagger}_i (\nu_i^{(1)})]
\end{equation}
${\tilde h}^{(1)}_i(\nu^{(1)}_i)$'s are the three point QED interaction vertices  and $\nu^{(1)}_i$ is the light-front energy transferred at the vertex ${\tilde h}^{(1)}_i$. 
$d\nu^{(1)}_i$ is the integration measure. For example, 
\begin{equation}
\int d\nu^{(1)}_1=\frac{1}{{(2\pi)}^{3/2}}\int{{[dp][dk]}\over{\sqrt{2\overline p^+}}} \;
\end{equation}
and  $\nu_1^{(1)}=p^--k^--(p-k)^-$. 
The expressions for $V_2(x^+)$ and $V_3(x^+)$  can be found in Ref.~\cite{JAI12}. 
Following the KF method,  $H_{as}$ is evaluated by taking the limit $x^+\rightarrow \infty$ in $exp[-i\nu_i^{(1)}x^+]$, which contains the time dependence of this term in the interaction Hamiltonian $H_{int}$. If $\nu_i^{(1)} \rightarrow 0$ for some vertex, then the corresponding term in $H_{int}$ does not vanish in large $x^+$ limit. One can then use KF method to obtain the asymptotic Hamiltonian and to construct the asymptotic M\"oller operator which leads to the coherent states.

Here, I will illustrate the construction of asymptotic Hamiltonian using the 3-point vertex only. As seen from the light-cone time dependence of $V_1(x^+)$,
 the non-zero contribution to asymptotic interaction Hamiltonian comes from the regions in which $\nu_i^{(1)}$'s vanish. 
It has been shown \cite{ANU94} that out of the four light-cone energy differences,  $\nu_i^{(1)}$'s  in Eq. (\ref{V_1as}), two can never  be zero. A convenient way to define the asymptotic region is by requiring 
\begin{displaymath}
{\bf k}_\perp ^2 < {{k^+ \Delta} \over{p^+}} \quad k^+ < {{p^+ \Delta} \over {m^2}}\;.
\end{displaymath}
where $\Delta = p^+\Delta E$. $\Delta E$ is an energy cutoff which may be choosen to be the experimental resolution. 
One can verify that in this region $\nu_i^{(1)}= p^--k^--(p-k)^- \rightarrow 0$. Therefore,  we can define 
$V_{1as}(x^+)$ as 
\begin{align}
V_{1as}(x^+)=e\sum_{i=1,4}\int d\nu_i^{(1)}\Theta_\Delta(k)[e^{-i \nu_i^{(1)} x^+}\tilde h_i^{(1)}(\nu_i^{(1)})+ e^{i \nu_i^{(1)} x^+}\tilde h^ {(1)\dagger}_i (\nu_i^{(1)})] \;
\end{align}
where
$\Theta_\Delta(k)$ is given by  
$$ \Theta_\Delta(k)=\theta\bigg({{k^+\Delta} \over p^+} - {\bf k}_\perp^2\bigg)
\theta\bigg({{p^+\Delta} \over m^2} - k^+\bigg)
$$\\
Taking the limit, $k^+\rightarrow 0$, ${\bf k_\perp} \rightarrow 0 $, in all slowly varying functions of k. Performing the $x^+$ integration and neglecting the 4-point instantaneous term, one obtains the asymptotic M\"oller operator $\Omega_{\pm}^A$ which gives the asymptotic states
\begin{align}
\Omega_{\pm}^A \vert n\colon p_i \rangle=exp\biggl[-e\int{dp^+d^2{\bf p}_\perp}\sum_{\lambda=1,2} [d^3k][f(k,\lambda:p)a^\dagger(k,\lambda)-f^*(k,\lambda:p)a(k,\lambda)]\biggr]\rho(p)\vert n \colon p_i \rangle
\end{align}
where \begin{equation}\label{theta2}
f(k,\lambda \colon p) = {{p_\mu\epsilon_\lambda^\mu(k)} \over {p\cdot k}}\theta\bigg(\frac{k^+\Delta}{p^+}-{\bf k}_\perp^2\bigg) \theta\bigg(\frac{p^+\Delta}{m^2}-k^+\bigg) \;,
\end{equation}
Taking into account, the asymptotic limit of instantaneous interaction also, one obtains \cite{JAI12}
\begin{align}
\Omega_{\pm}^A \vert  n\colon p_i \rangle=&exp\biggl[-e\int{dp^+d^2{\bf p}_\perp}\sum_{\lambda=1,2} [d^3k][f(k,\lambda:p) 
a^\dagger(k,\lambda)-f^*(k,\lambda:p)a(k,\lambda)]\nonumber\\
&+e^2\int{dp^+d^2{\bf p}_\perp}\sum_{\lambda_1,\lambda_2}
[d^3k_1][d^3k_2][g_1(k_1,k_2,\lambda_1,\lambda_2 \colon p) a^\dagger(k_2,\lambda_2)a(k_1,\lambda_1)\nonumber\\
&-g_2(k_1,k_2,\lambda_1,\lambda_2 \colon p)a(k_2,\lambda_2)a^\dagger(k_1,\lambda_1)]\biggr]\rho(p)\vert n \colon p_i \rangle
\end{align}
where 
\begin{align}
g_1(k_1,k_2,\lambda_1,\lambda_2 \colon p)=&-\frac{4p^+}{p \cdot k_1-p \cdot k_2+k_1 \cdot k_2} \delta^{3}(k_1-k_2)\nonumber\\
g_2(k_1,k_2,\lambda_1,\lambda_2 \colon p)=&\frac{4p^+}{p \cdot k_1-p \cdot k_2-k_1 \cdot k_2}\delta^{3}(k_1-k_2)
\end{align}

We have used light-cone time ordered perturbation theory to calculate the transition matrix elements in both Fock basis and the coherent state basis. 
The transition matrix is given by the perturbative expansion 
\begin{displaymath}
T= V + V {1 \over {p^--H_0}}V + \cdots
\end{displaymath}
Conventionally, electron mass shift $\delta m^2$  is obtained by calculating the matrix element of this series, $T_{pp}$, 
between the initial and final one electron  electron  Fock states $\vert p,s \rangle$:  
\begin{align}\label{deltam}
\delta m^2 = p^+ \sum_{s}  T_{pp}
\end{align}
Expanding $T_{pp}$ in powers of $e^2$ as 
\begin{align}
T_{pp}=T^{(1)}+T^{(2)}+\cdots
\end{align}
one gets $T^{(n)}$, the $O(e^{2n})$ contribution to fermion self energy correction. 

\section{Mass renormalization up to $O(e^2)$}
In $O(e^{2})$, fermion self energy correction is represented by the diagrams shown in Fig.~1. Fig.~1(b) is a tree level diagram and does not have any vanishing denominator. Neglecting the contribution of Fig.~1(b) 
\begin{figure}[h]
\includegraphics[scale=0.6]{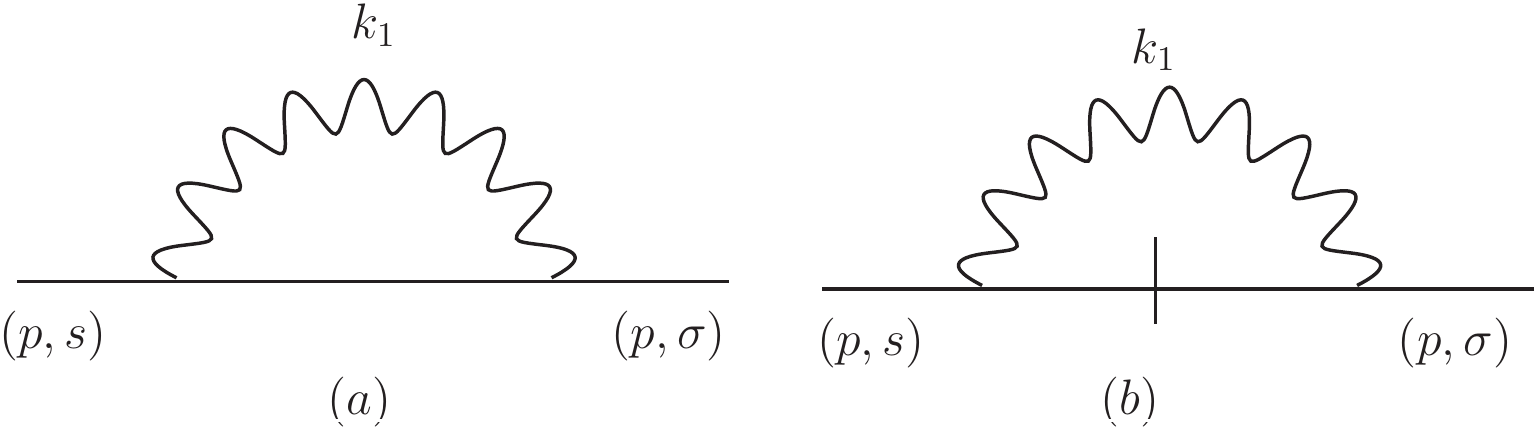}
\label{fig1}
\caption{Diagrams for $O(e^2)$ self energy correction in fock basis corresponding to $T_1$ in LF gauge}
\end{figure}
\begin{align}
T^{(1)}_{pp}\equiv T^{(1)}( p,p)=\langle p,s \vert V_1 \frac{1} {p^- - H_0}V_1\vert p,s \rangle
\end{align}
Following the standard procedure in Fock basis, we obtain $\delta m^2$ \cite{JAI12}, 
\begin{equation}\label{foc1}
\boxed{
	{(\delta m^2_{1a})}^{IR}=-\frac{e^2}{(2\pi)^3}\int{d^2{\bf k}_{1\perp}}\int\frac{dk_1^+}{k_1^+}\frac{(p\cdot \epsilon(k_1))^2}{(p\cdot k_1)}
}
\end{equation}
We have performed this calculation in Feynman gauge also. QED Lagrangian in Feynman gauge with additional PV fields is given by \cite{HILLER11}: 
\begin{align}  
{\cal L}= &\sum_{i=0}^2 (-1)^i \left[-\frac14 F_i^{\mu \nu} F_{i,\mu \nu} 
         +\frac{1}{2} \mu_i^2 A_i^\mu A_{i\mu} 
         -\frac{1}{2} \left(\partial^\mu A_{i\mu}\right)^2\right] \nonumber\\
& + \sum_{i=0}^2 (-1)^i \bar{\psi_i} (i \gamma^\mu \partial_\mu - m_i) \psi_i 
  - e \bar{\psi}\gamma^\mu \psi A_\mu.
\end{align}
\begin{figure}[h]
\includegraphics[scale=0.6]{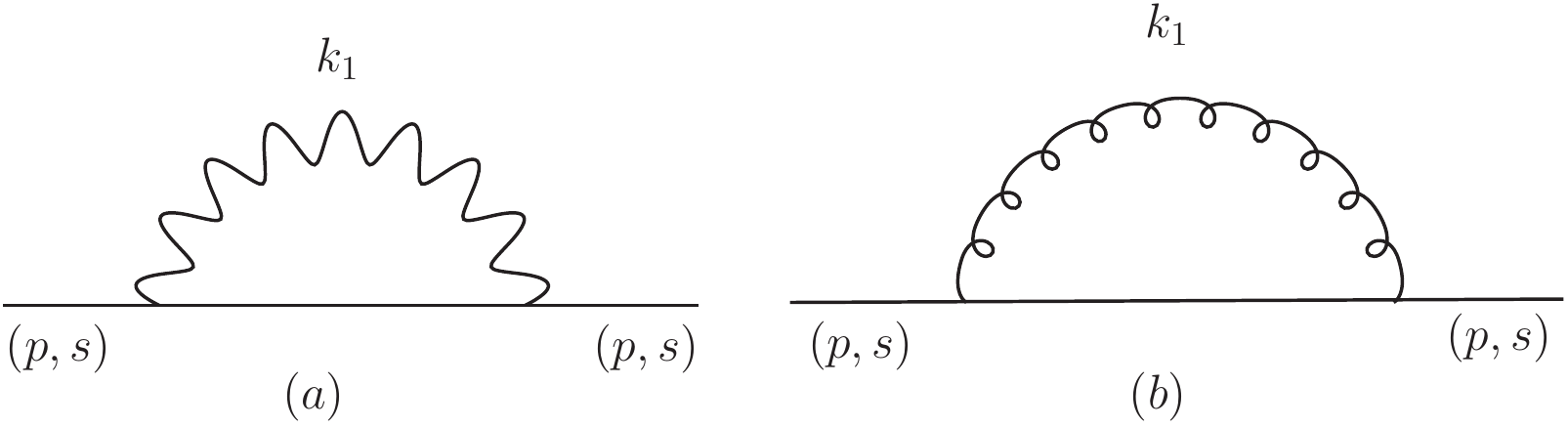}
\caption{Diagrams for $O(e^2)$ self energy correction in fock basis in Feynman gauge. In (a), wavy line corresponds to physical photon $(i=0)$ and in (b) curly line corresponds to PV photon $(i=1,2)$.}
\end{figure}

Here, there are additional contributions to self energy correction due to diagram in Fig.~2(b), in which the curly line denotes the massive PV field. Note that there are no instantaneous diagrams in Feynman gauge as the non-local terms in the Hamiltonian get canceled by similar terms for PV fields. Moreover, for massive PV photons there is no vanishing denominator. Thus, the diagram in Fig.~2(b) does not contribute to IR divergences. To conclude, up to one loop order, there is only one diagram that can give IR divergences in both LF and Feynman gauge.\\

\begin{figure}[h]
\includegraphics[scale=0.5]{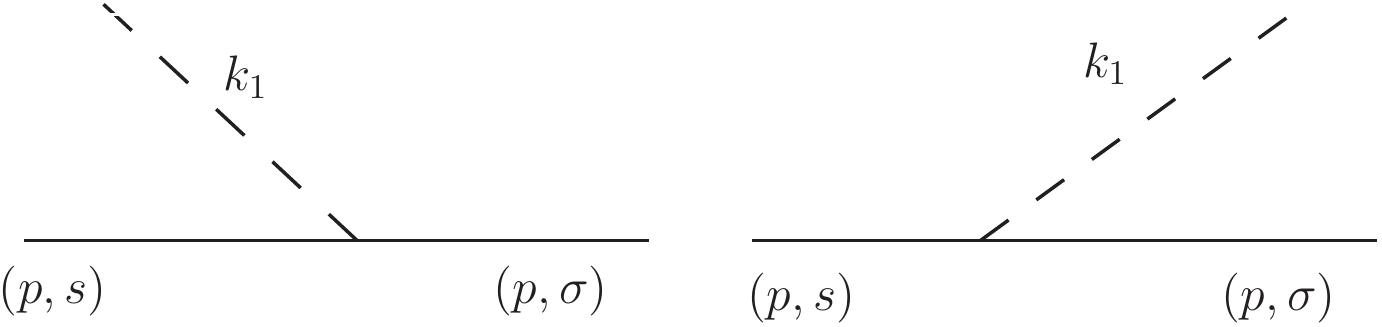}
\caption{Additional diagrams in coherent state basis for $O(e^2)$ self energy correction}
\end{figure} 
When we use coherent state basis to calculate the transition matrix elements, there are additional contributions apart from those already calculated in Fock basis, because the coherent state in the matrix element 
\begin{equation}
T^\prime(p,p)=\langle p,s \colon f(p) \vert V_1\vert p,s \colon f(p) \rangle
\end{equation}
contains $O(e)$ terms. In particular, in $O(e^2)$, one can also get a contribution from diagrams in Fig.~3(a), which represents a situation  where a soft photon accompanying the incoming particle is absorbed and Fig.~3(b) which represents a situation where a soft photon is emitted. However, the two particle states are indistinguishable from the single particle state due to finite experimental resolution.
The contribution of these coherent state diagrams is found to be 
\begin{align}
T^\prime(p,p)=&\frac{e^2}{(2\pi)^{3}}\int\frac{d^2{\bf k}_{1\perp}}{2p^+}\int \frac{dk_1^+}{2k_1^+}\nonumber\\
&\times \overline u(\overline p,s^\prime)\epsilon\llap/^\lambda(k_1)u(p,s)f(k_1,\lambda:p)
\end{align}
where the form of  $f(k,\lambda \colon p)$ in Eq.~\ref{theta2} ensures that the integrals are performed only over a small region around $k ^+= 0$, ${\bf k_\perp} = 0 $. Using Eq.~(\ref{deltam}) and simplifying further we obtain 
\begin{align}\label{coh1}
\boxed{
  {(\delta m^2)}^\prime=\frac{e^2}{(2 \pi)^3} \int{d^2{\bf k}_{1\perp}}\int\frac{dk_1^+}{k_1^+}\frac{(p\cdot \epsilon(k_1))^2 \Theta_\Delta (k_1)}{p\cdot k_1}
}
\end{align}
Adding the IR divergent contributions arising due to $p.k_1 \rightarrow 0$ in Eqs. (\ref{foc1}) and (\ref{coh1}), we find that IR divergence are canceled in the coherent state basis in  $O(e^2)$.
\section{Mass renormalization up to $O(e^4)$}
In $O(e^4)$, the contribution to self energy correction in the Fock basis comes from diagrams which contain
\begin{itemize}
\item only 3-point vertices which is represented by Fig.~4.
\item both 3-point and 4-point vertices which is shown in Fig.~5.
\item only 4-point vertices which is represented by Fig~6.
\end{itemize}
\begin{figure}[h]
\centering
\includegraphics[scale=0.5]{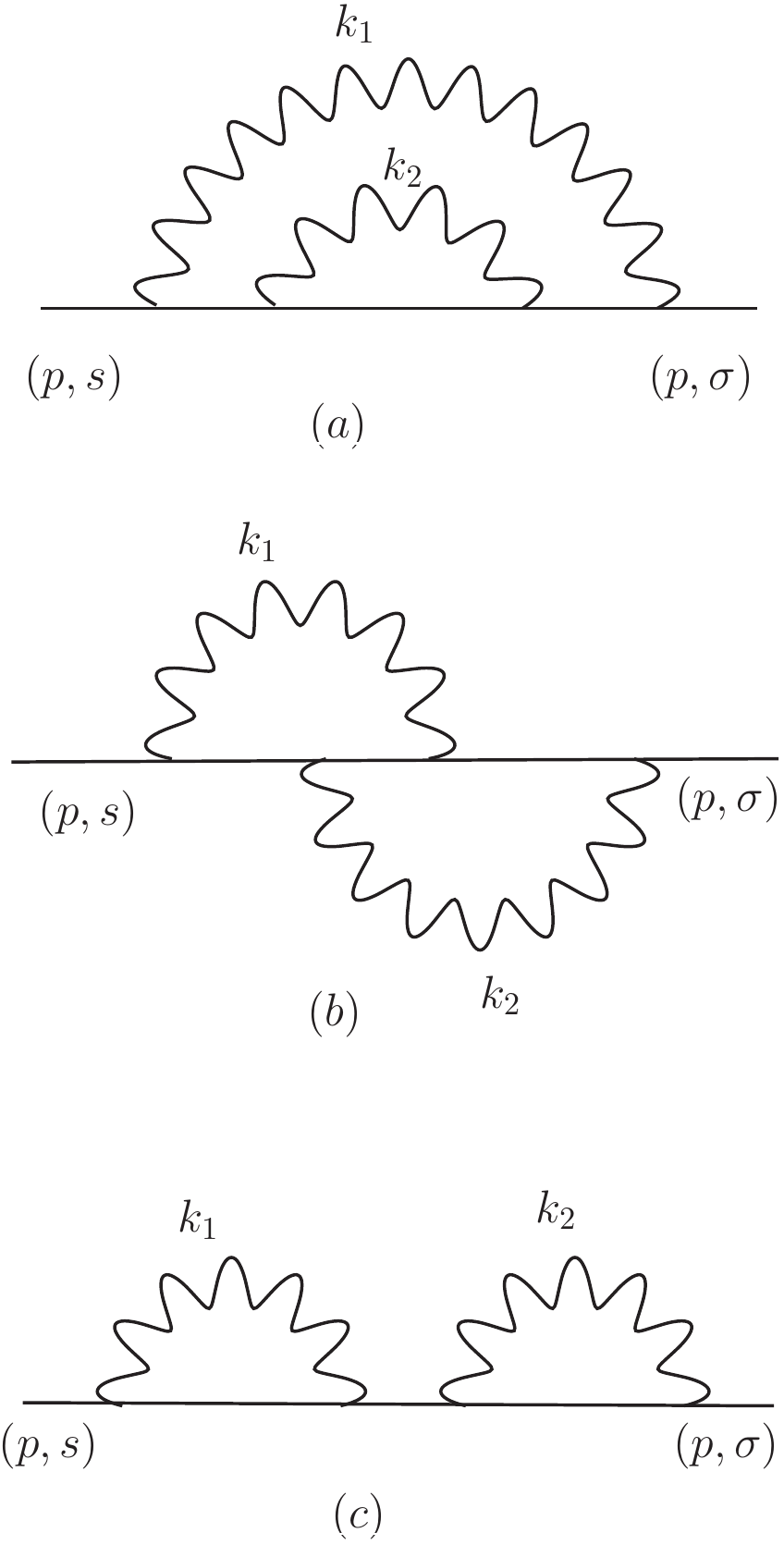}
\label{fig4}
\caption{Diagrams for $O(e^4)$ self energy correction corresponding to $T_3$ in fock basis in LF gauge}
\end{figure}
\begin{figure}[h]
\centering
\includegraphics[scale=0.4]{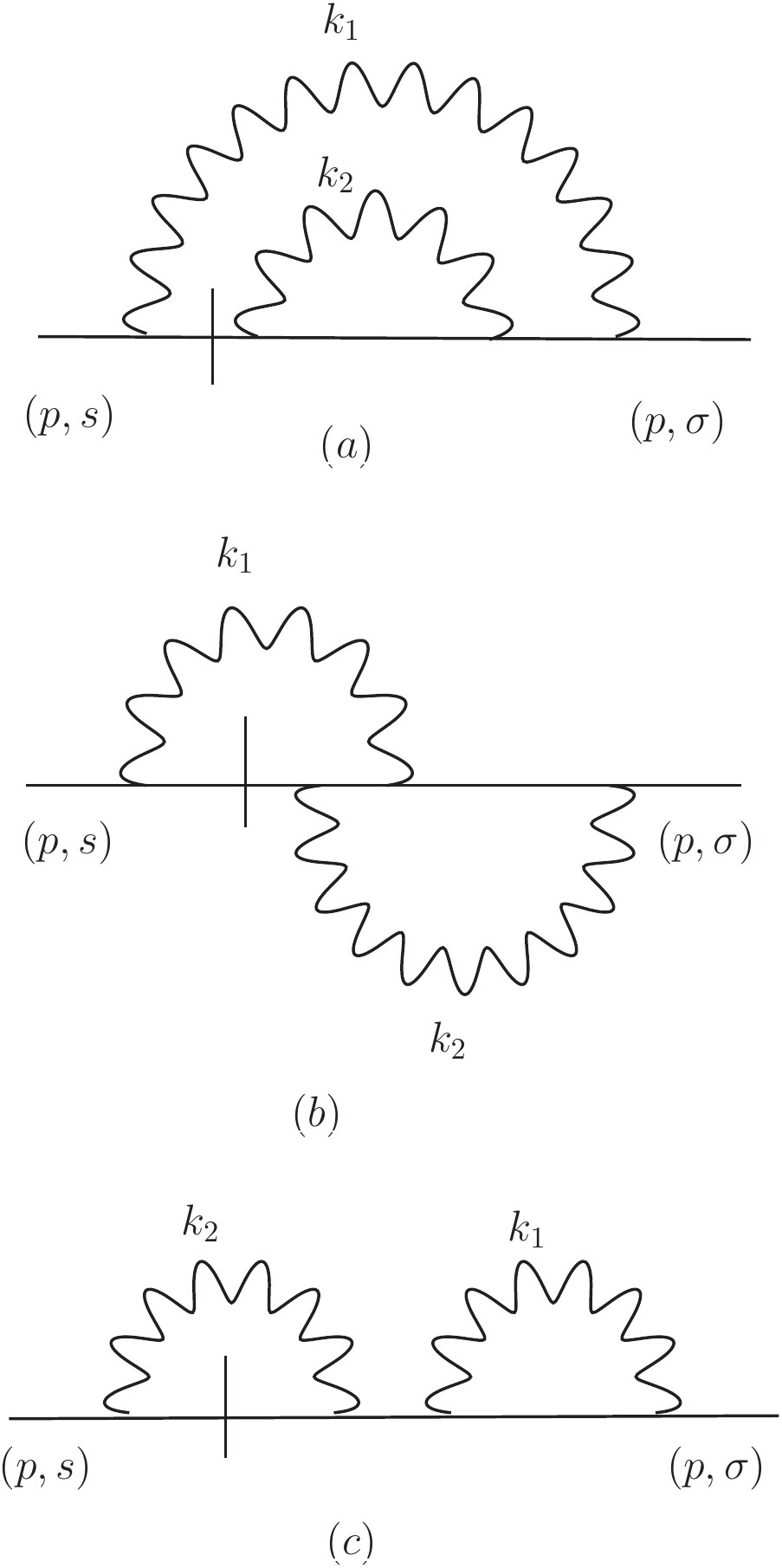}\includegraphics[scale=0.35]{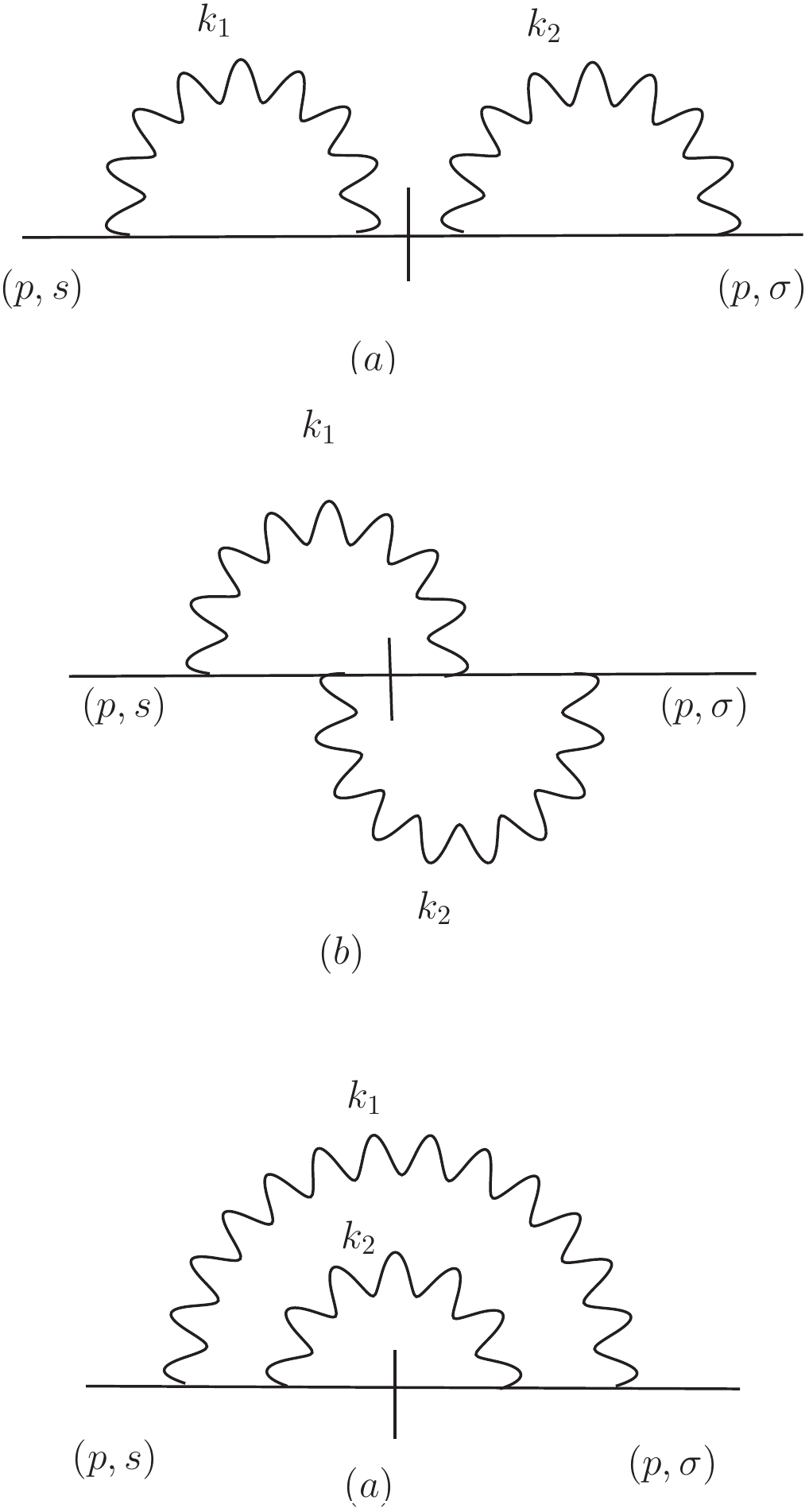}\includegraphics[scale=0.4]{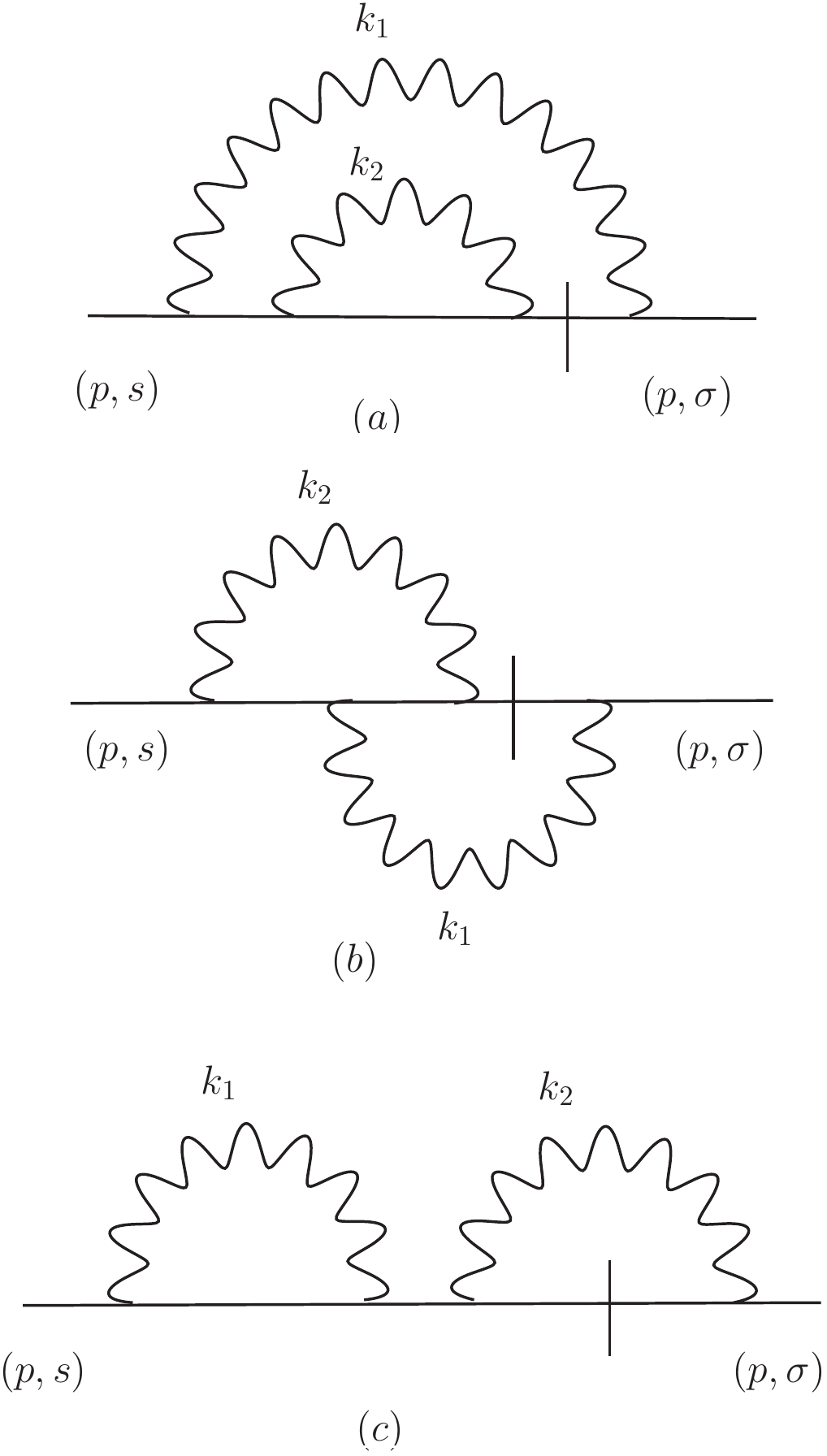}
\caption{Diagrams for $O(e^4)$ self energy correction in fock basis corresponding to $T_4$,  $T_5$ and $T_6$ respectively.}
\end{figure}
\begin{figure}[h]
\centering
\includegraphics[scale=0.4]{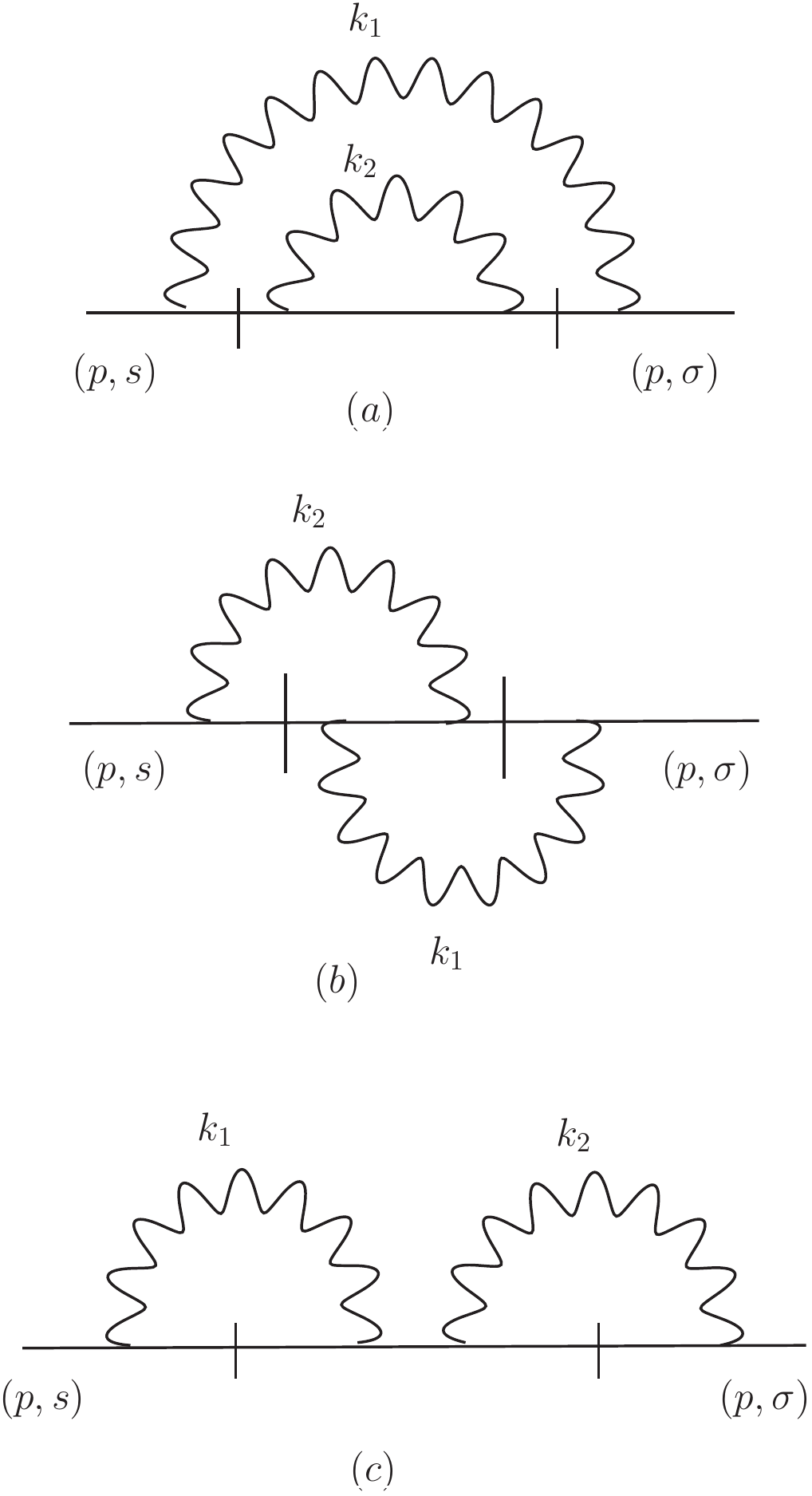}
\caption{Diagrams for $O(e^4)$ self energy correction in fock basis corresponding to $T_7$,}
\end{figure}
Transition matrix element for $O(e^4)$ correction to self energy is given by
\begin{displaymath}
T^{(2)}=T_{3}+T_{4}+T_{5}+T_{6}+T_{7}                       
\end{displaymath}
where
\begin{align}
T_{3}=&\langle p,s \vert V_1 \frac{1} {p^- - H_0}V_1 \frac{1} {p^- - H_0}V_1 \frac{1} {p^- - H_0}V_1 \vert p,s \rangle \\ \label{T_3}
T_{4}=&\langle p,s \vert V_1\frac{1} {p^- - H_0}V_1\frac{1} {p^- - H_0}V_2\vert p,s \rangle\\ \label{T_5}
T_{5}=&\langle p,s \vert V_1\frac{1} {p^- - H_0}V_2\frac{1} {p^- - H_0}V_1\vert p,s \rangle\\ \label{T_6}
T_{6}=&\langle p,s \vert V_2\frac{1} {p^- - H_0}V_1\frac{1} {p^- - H_0}V_1\vert p,s \rangle\\ \label{T_7} 
T_{7}=&\langle p,s \vert V_2\frac{1} {p^- - H_0}V_2\vert p,s \rangle
\end{align}
We have calculated these diagrams using light-cone time ordered perturbation theory and have shown that there are 
IR divergences present in the Fock basis in Figs.~4 and 5. \\
Transition matrix element for $O(e^4)$ correction to self energy due to the additional contribution in coherent state  basis is given by
\begin{figure}[h]
\centering
\includegraphics[scale=0.4]{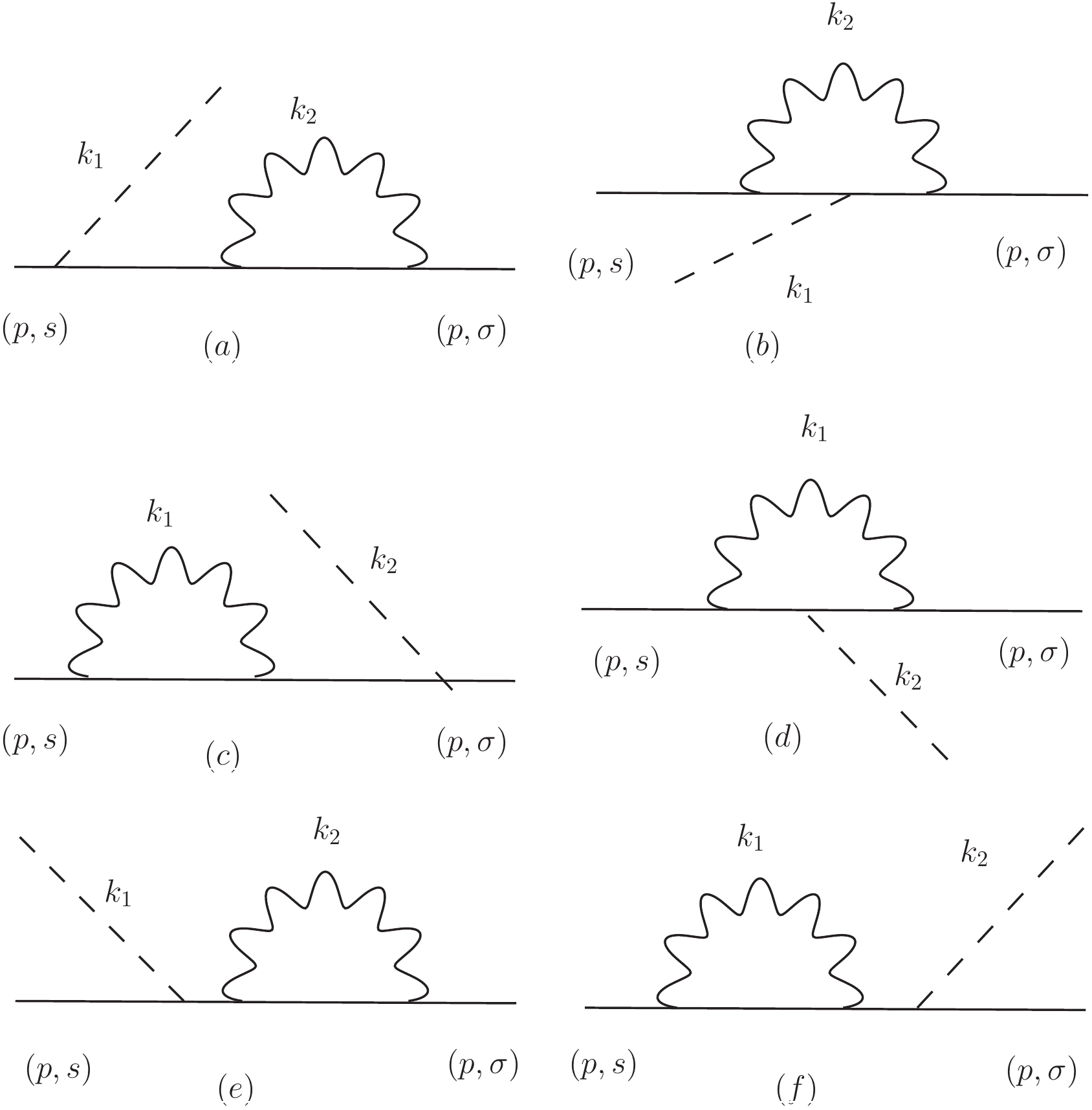}
\caption{Additional diagrams in coherent state basis for $O(e^4)$ self energy correction corresponding to $T_8$.}
\end{figure}
\begin{figure}[h]
\centering
\includegraphics[scale=0.4]{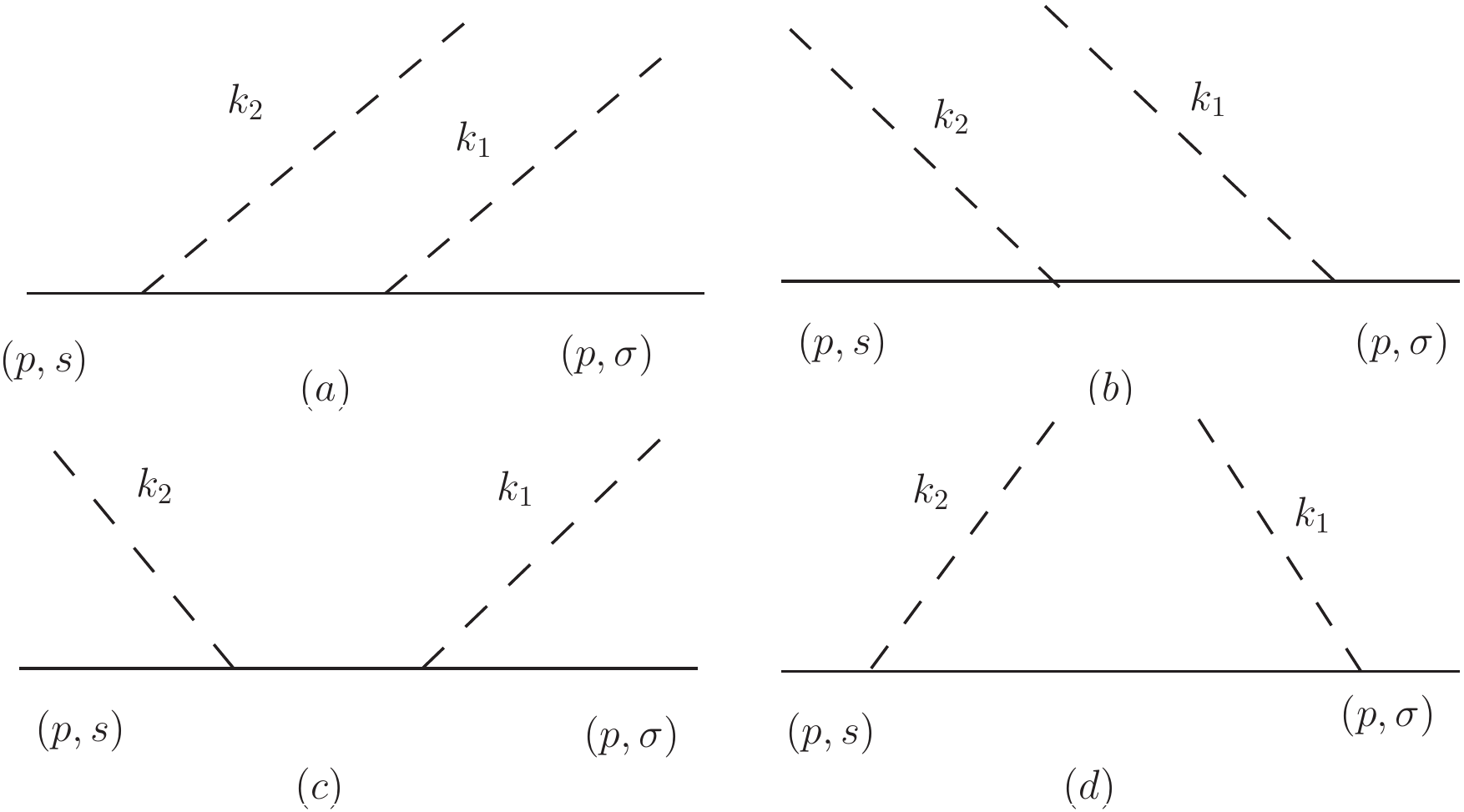}
\caption{Additional diagrams in coherent state basis for $O(e^4)$ self energy correction corresponding
 to $T_9$.}
\end{figure}
\begin{displaymath}
T^{(2)}+T_8^\prime+T_9^\prime+T_{10}^\prime+T_{11}^\prime \,
\end{displaymath}
where 
\begin{align}
T_8^\prime=&\langle p,s \colon f(p) \vert V_1 \frac{1} {p^- - H_0}V_1\frac{1} {p^- - H_0}V_1\vert p,s \colon f(p) \rangle,\\
T_9^\prime=&\langle p,s \colon f(p) \vert V_1 \frac{1} {p^- - H_0}V_1\vert p,s \colon f(p) \rangle, \\
T_{10}^\prime=&\langle p,s\colon f(p)\vert V_1\frac{1} {p^- - H_0}V_2\vert p,s \colon f(p)\rangle+\langle p,s\colon f(p)\vert V_2 \frac{1} {p^- - H_0}V_1\vert p,s \colon f(p)\rangle \\
T_{11}^\prime=&\langle p,s\colon f(p)\vert V_2 \vert p,s \colon f(p)\rangle 
\end{align}
In Ref.~\cite{JAI12}, it was shown that in $O(e^4)$, the IR divergences get canceled if coherent state basis is used in LF gauge. In Section 2, we have verified this cancellation in Feynman gauge also up to $O(e^2)$ thereby establishing the usefulness of the coherent state basis in both LF and Feynman gauge. We have also verified that IR divergences cancel up to $O(e^4)$ in Feynman gauge. Details can be found in Ref.~\cite{JAI13}. 
\begin{figure}[h]
\centering
\includegraphics[scale=0.4]{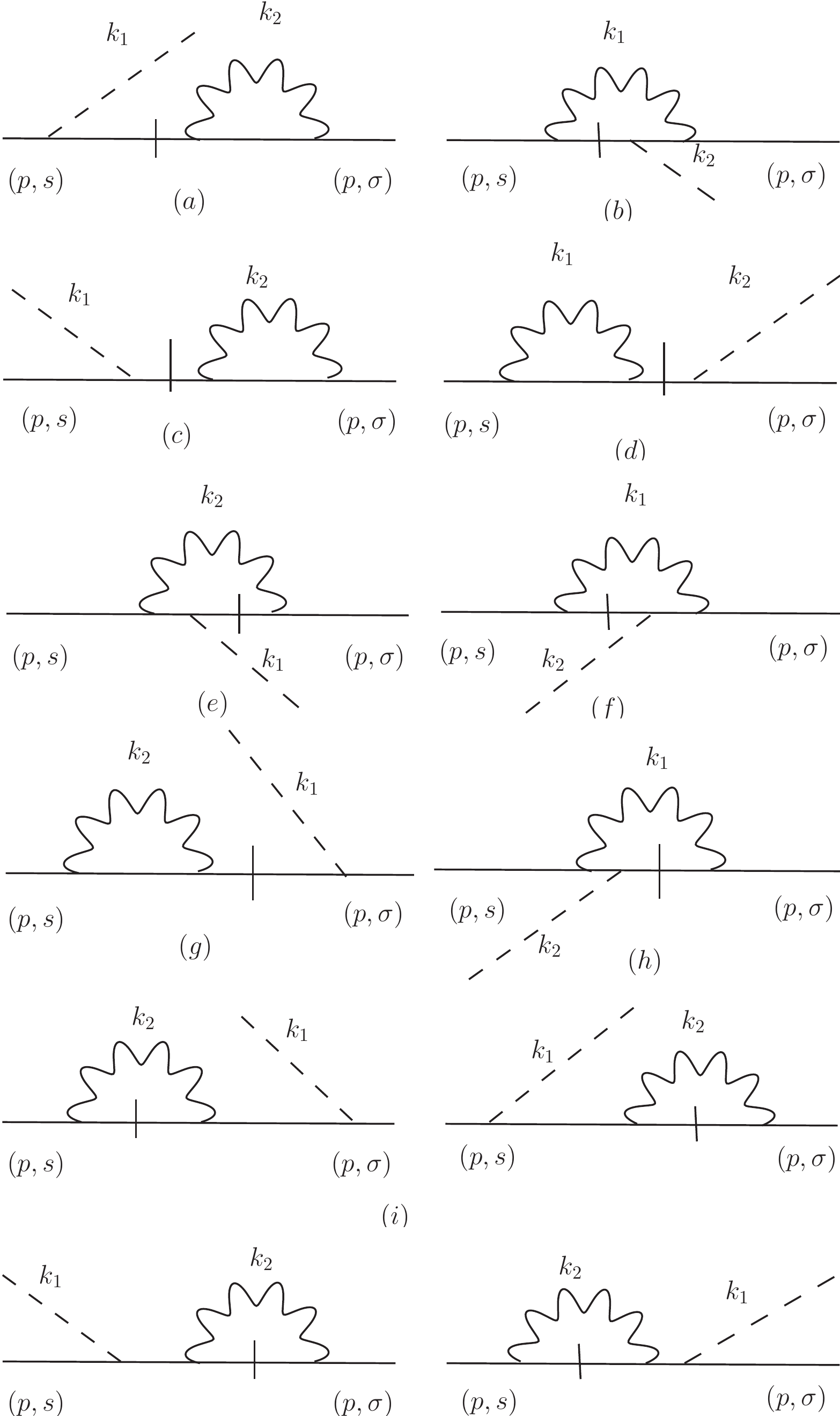}
\caption{Additional diagrams in coherent state basis for $O(e^4)$ self energy correction corresponding to $T_{10}$.}
\end{figure}
\begin{figure}[h]
\centering
\includegraphics[scale=0.4]{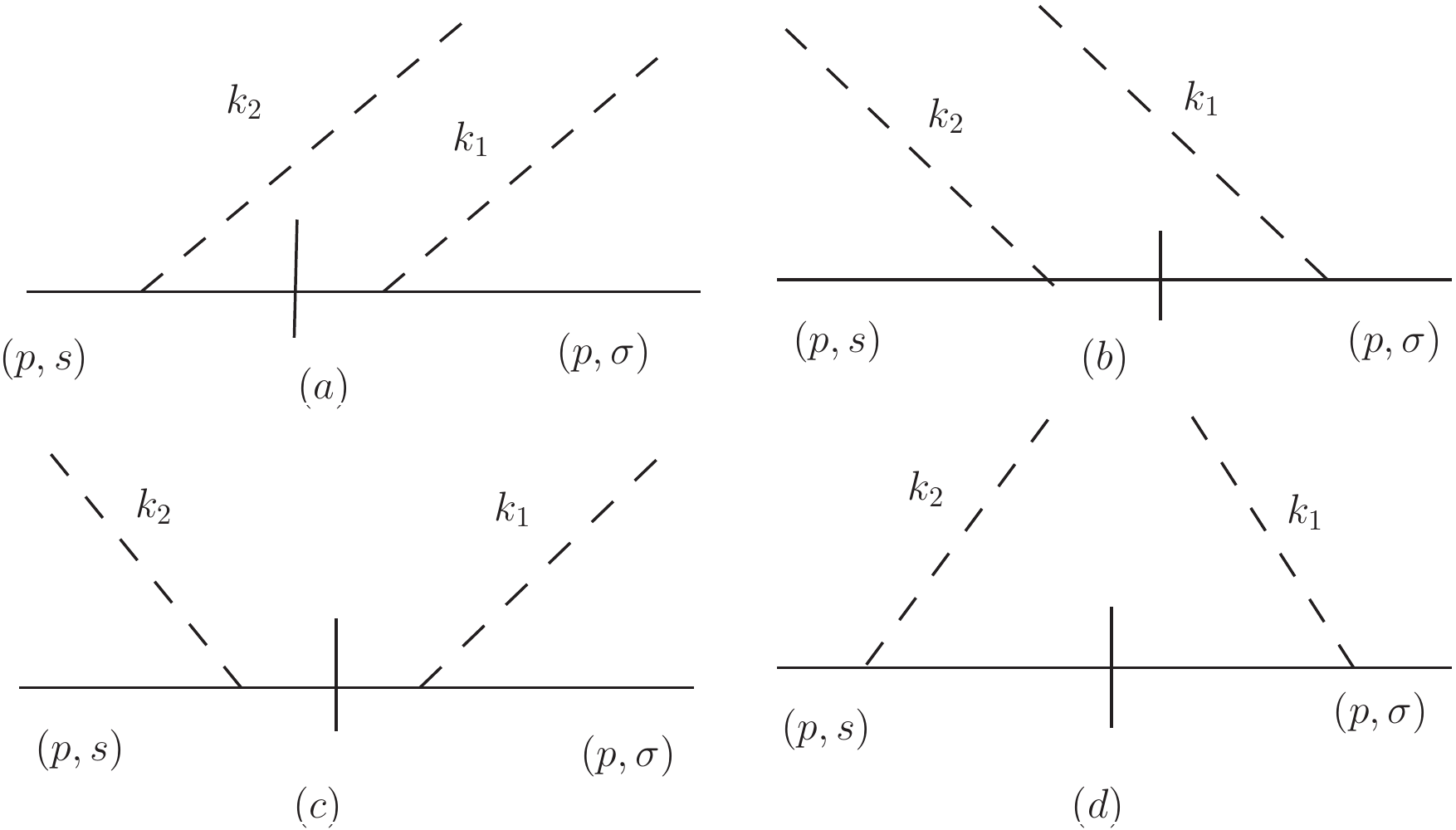}
\caption{Additional diagrams in coherent state basis for $O(e^4)$ self energy correction corresponding to $T_{11}$.}
\end{figure}
\section{Conclusion}
We have shown that the true IR divergences get canceled up to $O(e^4)$ when coherent state basis is used to calculate the transition matrix elements in lepton self energy calculation in light-front QED in LF gauge as well as in Feynman gauge.
The cancellation of IR divergences between real and virtual processes is known to hold 
in equal time QED to all orders. It would be interesting to verify this all order cancellation in LFQED. 
The present work is an initial step in this direction. It is well known that IR divergences do not cancel in QCD in higher orders. This is related to the fact that the asymptotic states are bound states. Connection between asymptotic dynamics and IR divergences can possibly be exploited to construct an artificial potential that may be used in bound state calculations in LFQCD.
\vskip 0.5cm
{\bf ACKNOWLEDGEMENTS}
\vskip 0.25cm
I would like to thank University of Delhi and organizers of LC2012 for travel support. Part of this work was done under a DST sponsored project No. SR/S2/HEP-17/2006.

\end{document}